\documentclass[conference]{IEEEconf}
\input epsf
\usepackage{graphicx}

\usepackage{amsmath}
\usepackage{amssymb}
\usepackage{xcolor}
\usepackage{amsthm}
\usepackage{mathtools}
\usepackage{siunitx}
\usepackage{cuted}
\usepackage{graphicx}
\usepackage{subcaption}
\usepackage{epstopdf}
\usepackage{tabularx}
\usepackage{multirow}
\usepackage{booktabs}
\usepackage{diagbox}
\usepackage{cuted}
\usepackage{lipsum}
\usepackage[noadjust]{cite}

\usepackage[english]{babel}
\usepackage[utf8]{inputenc}
\usepackage{algorithm}
\usepackage{algpseudocode}

\usepackage{geometry}

\usepackage{optidef}

\makeatletter
\newcommand{\removelatexerror} {\let\@latex@error\@gobble}
\makeatother

\newif\iftrackrivision

\iftrackrivision

\else

\fi

\begin{document}
	


\title{\textbf{\Large Flexible and dependable manufacturing beyond xURLLC: A novel framework for communication-control co-design\\}}

\author{Bin~Han$^{1,*}$, ~Mu-Xia~Sun$^{2}$, ~Lai-Kan~Muk$^{1}$, ~Yan-Fu~Li$^{2}$ and ~Hans~D.~Schotten$^{1}$\\
	\normalsize $^{1}$Technische Universit\"at Kaiserslautern, Kaiserslautern, Germany\\
	\normalsize $^{2}$Tsinghua University, Beijing, China\\
	\normalsize bin.han@eit.uni-kl.de, muxiasun@tsinghua.edu.cn, laimuk@rhrk.uni-kl.de, liyanfu@tsinghua.edu.cn, schotten@eit.uni-kl.de\\
	\normalsize *corresponding author
}

	\newgeometry{top=72pt, left=54pt, bottom=54pt, right=54pt}
	
	\maketitle
	
	\begin{abstract}
		Future Industrial 4.0 applications in the 6G era is calling for high dependability that goes far beyond the current ultra-reliable low latency communication (URLLC), and therewith proposed critical challenges to the communication technology. Instead of struggling against the physical and technical limits towards an extreme URLLC (xURLLC), communication-control co-design (CoCoCo) appears a more promising solution. This work proposes a novel framework of CoCoCo, which is not only enhancing the dependability of 6G industrial applications such as remote control, but also exhibiting rich potential in revolutionizing the future industry per openness and flexibility of manufacturing systems.
	\end{abstract}
	
	
	\IEEEpeerreviewmaketitle

	\section{Introduction}\label{sec:intro}
	A characterizing feature of the Fifth Generation (5G) mobile communication systems, ultra-reliable low-latency communication (URLLC), is the first use case of wireless communication that has ever been proposed to set strict constraints on both latency and reliability. Being an emerging concept, URLLC has been under rapid evolution and its requirements remains being frequently updated in the 5G standards. For instance, the two URLLC scenarios with most extreme quality of service (QoS) requirements defined by the 3\textsuperscript{rd} Generation Partnership Project (3GPP) in their Release 15 (Rel.15) are \emph{process automation - remote control} and \emph{electricity distribution - high voltage}, which require 99.9999\% (``six nines'') reliability within \SI{60}{\milli\second} end-to-end (E2E) latency and ''five nines'' reliability within \SI{5}{\milli\second} E2E latency, respectively~\cite{3gpp22261f90}. However, in the upcoming Rel. 16, these frontiers will be further pushed to ``six nines'' in \SI{1}{\milli\second} and ``five nines'' in \SI{0.5}{\milli\second}, by two newly introduced scenarios of \emph{discrete automation – motion control} and \emph{tactile interaction}, respectively~\cite{3gpp22261gg0}. Looking forward to the beyond-5G (B5G) and Sixth Generation (6G) technologies, even more extreme requirements are envisaged for the so-called extreme URLLC (xURLLC) scenario. For example, the requirements of ``seven nines'' reliability within sub-\SI{1}{\milli\second} E2E latency and ``nine nines'' reliability within \SI{0.1}{\milli\second} E2E latency have been proposed in \cite{PSS+2020}, to serve the applications of wireless factory automation and Internet of Sensors, respectively.
	
	Such stringent QoS requirements on E2E latency and communication reliability will inevitably raise several tremendous technical challenges, which are not likely to be resolved by any conventional technology at a reasonable cost. As a potential way out of this dilemma, the co-design of control system and communication system in context of URLLC applications has raised an extensive interest over the past years. Significant investigation efforts have been put into this emerging research field to answer the question posed in \cite{PSS+2020}: \emph{Can URLLC requirements be relaxed by taking into account control dynamics, while ensuring control stability?} Despite of some initial achievements, yet there has been no systematic methodology developed to solve the generic problem of communication-control co-design (CoCoCo).
	
	In this paper, we study the CoCoCo problem in a system engineering perspective. By shedding the light on the sources, types, propagation, and impact of different errors in networked control systems (NCS), we set up a generic abstract model for such systems. Based on this model, we propose a novel framework to co-design and co-optimize the modules of communication, control, and computation in NCS, which not only can address the xURLLC challenges, but also may revolutionize the future industry paradigm by enabling open and flexible manufacturing systems.
	
	The remainder of this paper is organized as follows: We start with Sec.~\ref{sec:challenges} to identify the technical challenges in strictly fulfilling the xURLLC performance requirements. Then in Sec.~\ref{sec:dependability}, we explain the 6G philosophy of preferring the application dependability over the communication reliability which is emphasized in 5G URLLC. 
	Afterwards, we present in Sec.~\ref{sec:approaches} our proposed approaches of CoCoCo analysis, which consists of a multi-domain NCS model, a decomposition of this model into domain-specific modules that are coupled to each other, and a discussion on identifying the domain-coupling metrics. Afterwards, with Sec.~\ref{sec:applications} we set forth the application scenarios of our proposed methodology, revealing how it shall impact the future industry in various aspects, before concluding our work and closing this paper in Sec.~\ref{sec:conclusion}.
	
	\newgeometry{top=54pt, left=54pt, bottom=54pt, right=54pt}
	
	\section{Technical dilemma of vanilla xURLLC}\label{sec:challenges}
	On the one hand, the E2E latency in wireless systems consists of multiple components, namely the over-the-air transmission delay, the queuing delay, the processing/computing delay, and extra delay caused by re-transmissions. Any increase in one or multiple among these delays will reflect in an increased E2E latency.
	
	On the other hand, to increase the data-link reliability, there are generally four options: \emph{i}) to improve the SINR; \emph{ii}) to apply higher bandwidth; \emph{iii}) to use lower coding rate; and \emph{iv}) to leverage retransmission schemes such as automatic repeat request (ARQ) and hybrid HARQ (HARQ). 
	
	Unfortunately, the gain that can be achieved with option \emph{i}) is very limited, due to the regulations of maximal transmission power and the logarithmic increase of channel capacity upon SINR. Option \emph{ii}) can be effective for a single radio link, however, in large-scale networked systems where numerous devices are sharing the same radio resource pool, a higher bandwidth for every single device implies a lower multiplexing rate, and eventually a higher queuing delay. A lower coding rate suggested in option \emph{iii}) will certainly help reduce the packet error rate with its extra redundancy, however, it also leads to a longer message length, i.e. higher over-the-air transmission delay; additionally, it is also requiring more complex processes of coding and decoding, which may increase the computing delay. At last, every retransmission attempt will at least double the E2E latency w.r.t. a single transmission (usually taking even longer upon feedback latency and incremental redundancy). Therefore, the option \emph{iv}) is practically disabled from xURLLC applications.

	To summarize, the dilemma of vanilla xURLLC is originated from the conflicting requirements of E2E latency and link reliability, which fail to compromise each other in context of xURLLC. To the best of our knowledge, there has been no conventional approach that can promisingly resolve this challenge.
	
	\section{I4.0 in 6G: Dependability beyond URLLC}\label{sec:dependability}
	As an alternative to vanilla xURLLC, the emerging concept of dependability, which addresses reliability,
	availability, safety, integrity, and maintainability as a crucial prerequisite for mission-critical application~\cite{URB+2021}, is now attracting research interest in scope of 6G industrial communications. 
	
	\subsection{Networkd control systems and its dependability}
\begin{figure}[!htpb]
		\centering
		\includegraphics[width=.6\linewidth]{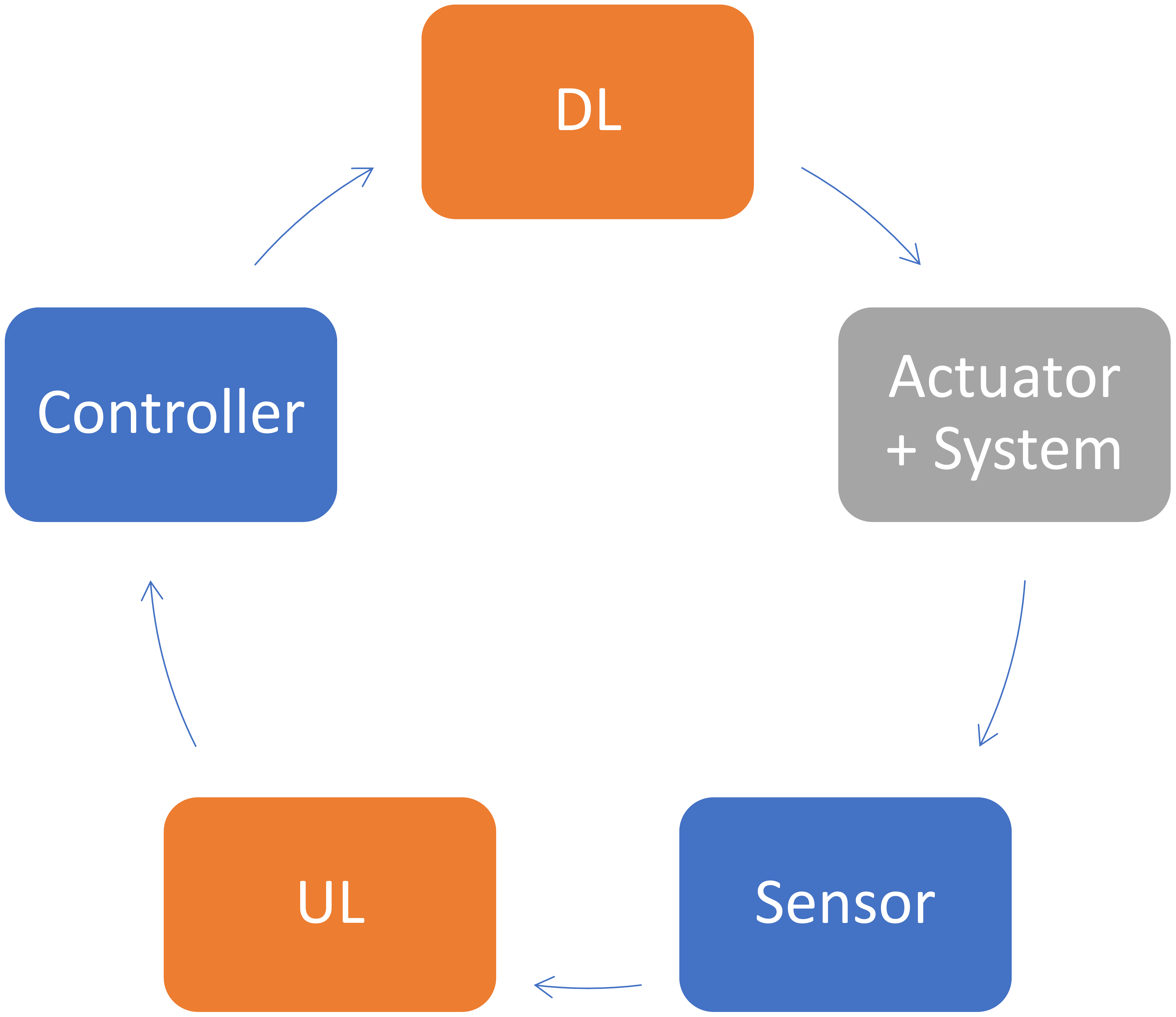}
		\caption{Abstracted model of networked control systems}
		\label{fig:ncs_model}
	\end{figure}

	Networked control systems, which are dominating the industrial use scenarios of 6G, can be generically modeled by Fig.~\ref{fig:ncs_model}. The controller receives feedback information about the instantaneous system status from sensor via an uplink (UL) channel, and correspondingly generates control commands, which it then sends over the downlink (DL) channel to the actuator. The actuator will then execute the commands, which eventually influences the system status that can be captured and measured by the sensor.
	
\begin{figure}[!htpb]
		\centering
		\includegraphics[width=.5\linewidth]{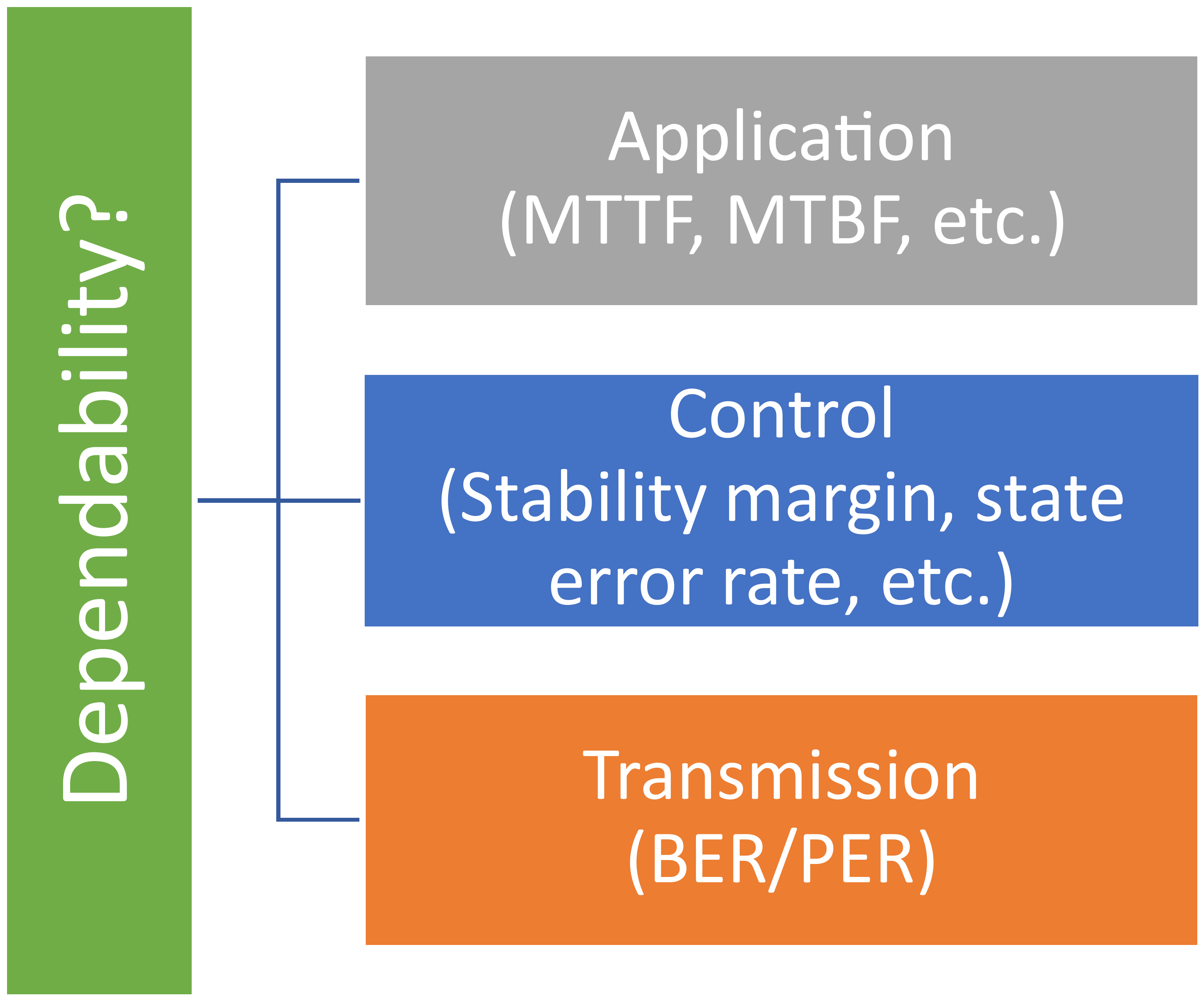}
		\caption{Dependability model of networked control systems}
		\label{fig:dependability_model}
	\end{figure}

	The dependability problem in such systems can be investigated in three different domains that lay over each other, as illustrated in Fig.~\ref{fig:dependability_model}. On top is the application domain regarding the industrial use cases, where the dependability is generally described by metrics related to application failures, such as the mean time till failure (MTTF), the mean time between failures (MTBF), etc. Beneath that is the control domain that describes the control system, where the dependability is usually evaluated by criteria such like the system stability margin, the system error regarding ideal state, etc. At the bottom is the telecommunication domain that deals with the information transmission over the UL and DL channels, where the dependability can be simplified into link reliability, which is generally evaluated by the transmission errors, e.g. bit error rate (BER) or packet error rate (PER).
 \subsection{Dependability of industrial systems}\label{subsec:depend_ind}
	Now we present a more detailed description of the system-level dependability for network-control-based systems from the view of reliability engineering (which is an important sub-discipline of system engineering that describes the ability of the system and its components to function without failure). Traditionally, industrial systems are formulated in binary-state reliability models: i.e., the system and each of its components has two possible working states: functioning or failure. The states are modeled as random variables, and there evolution are considered as stochastic or physical processes. Specifically, the state of the system is determined by its structure function 
\begin{eqnarray}
X_\text{sys,t}:=\Phi(X_{1,t},X_{2,t},...,X_{n,t})\
\end{eqnarray}
where $X_\text{sys,t}$ is the system reliability, each $X_{i,t}$ is the reliability of component $i$ at time $t$. Then, the system dependability metrics, such as MTTF \& MTTB, are evaluated through the probabilistic integrals over the process $X_\text{sys,t}$.

In recent years, the multi-state system models, which assume the system/components may have multiple working states between perfect functioning and failure, has also been wildly applied. However, all these standard system reliability models are insufficient for modelling the dependability of complex industrial system involving the CoCoCo challenges. Performance reliability, which measures the probability that the system to successfully perform the give control tasks, should also be used to incorporated with the standard system reliability models. To be more precise, a performance reliability is usually defined in the following form
\begin{eqnarray}
R_\text{per}(t)=\mathbb{P}(X_{\text{per},t} \in \mathcal{D}_\text{per} )\
\end{eqnarray}
\noindent where $R_\text{per}(t)$ is the performance reliability at time $t$, $X_{\text{per},t}$ is a random variable represents the performance at $t$, while $\mathcal{D}_\text{per}$ is the demand set, indicating the set of satisfying performance states.

Generally speaking, the QoS reliability metrics for communication networks, e.g., the probability of E2E  latency or data rates being less than certain threshold, are belongs to the type of performance reliability metrics. (For simplicity, such reliability metrics will be called as the E2E latency reliability or E2E data rate reliability, in the rest of the paper.) 

In the industrial sectors however, our concerns are more focused on the performance reliability of the actuators, at the application level. Unfortunately, to define, access, and evaluate such performance reliability could be a challenging task. It involves (possibly time-variant) stochastic time delays upon the industrial control systems, while these time delays are mappings of both the physical environments and the CoCoCo design parameters of the wireless communication  systems. As an essential task in CoCoCo for application-level design, we expect to estimate the following regression function: 
\begin{eqnarray}\label{1}
R_\text{sys,app}(t)=f(\textbf{\emph{R}}_\text{com}(t))\
\end{eqnarray}
\noindent where $R_\text{sys,app}(t)$ is the the system's performance reliability at the application level, while $\textbf{\emph{R}}_\text{com}(t)$ is the vector of communication network performance reliabilities, such as the E2E latency / RAN delay / data rate reliability.

 \subsection{Controller design considering system dependability}\label{subsec:depend_con}
 In the fields of control, the above system dependability problems are often handled in a different perspective. Generally, controllers are expected to be more robust against component failures or stochastic time-delays. It results in the designing of fault-tolerance controllers, reliable controllers and controllers for systems with stochastic time delays~\cite{MXZ+2019,HZ2022,HZS+2021,BR2005}. The designed controllers are often capable to handle the impact of stochastic E2E time-delays, when certain assumptions are satisfied. Still, optimal designs of controllers cannot fully guarantee a 100\% performance reliability at the application level. For instance, when congestion appears in the communication network and the E2E communication is interrupted. Thus, the optimal design of controllers must be considered as part of the system-level reliability design, while it should also be incorporated with the design of communication networks, which determines how the time-delays are injected into the control systems.

	\subsection{Error coupling and top-down design}\label{subsec:err_prop}
	This overlaying system architecture inevitably results in an error coupling through the three domains: transmissions errors will reduce the stability of control system, and eventually degrade the application dependability. Classically, a top-down design paradigm has been applied in industrial use scenarios: First, the entire system is first modeled and its dependability analyzed on the application layer. Regarding the system model, the optimal control shall be designed assuming an ideal communication, i.e. with consistent E2E latency and no error at all, to fulfill the application's requirement of dependability. Afterwards, this optimal control system must be analyzed regarding its tolerance to transmission errors, through which a strict requirement of communication reliability and E2E latency can be outlined. At last, it is the task of communication engineering to achieve the targeted performance level.
	
	Obviously, this methodology decomposes a multivariate optimization problem across three domains into several independent domain-wide sub-problems, which greatly simplifies the complexity of solving. However, this simplification is not supported by any mathematical structure or analytical feature (e.g. joint convexity or monotonicity), so it cannot guarantee to achieve the global optimum of the original problem. Instead, it is very likely ending up with a sub-optimum. 
	This claim is well supported by the evidence observed in the autonomous driving scenario studied in \cite{JFZ+2020}, where the optimal safe inter-vehicle distance is achieved at neither the control-preference (low status update interval) nor the communication-preference (vanilla URLLC), but a compromised specification between them.
	
	\subsection{Co-design as an emerging requirement of future I4.0}
	Over decades, this problem had not raised much research interest, because there had been no decent use case of remote control over wireless. In classical industrial systems, controllers, sensors, and actuators are typically embedded together and connected to each other by wired channels, which can be extremely reliable even at very high data rate. In this case, the communication domain fades out and the three-layer dependability model is simplified into a two-layer version with practically \emph{ideal} communication links, so the aforementioned problem does not exist.
	
	However, as modern and future industrial applications are dealing with more complex tasks, such as flexible manufacturing, collaborative robot control, etc., the demand to computing capability is explosively increasing. Meanwhile, the 6G ambition to connect \emph{everything} and \emph{everyone} is implying a higher integration level with compacter devices that have smaller or even no battery. Additionally, regarding the higher mobility of devices, increasing flexibility of industrial environment, and extremer working conditions of systems, wired connections will become gradually less applicable for inter-machine and machine-to-infrastructure communications in the future I4.0 scenario. Instead, MEC-based remote controlling will be essential, which brings the CoCoCo problem into our vision.
	%
	
	\section{From alternating optimization in-loop to modular co-design}\label{sec:approaches}
	\subsection{Application-oriented dependability management}\label{subsec:in-loop-design}
	The most straightforward solution to resolve the co-design problem introduced in the last section is a loopback design, where the control and communication components are optimized by turns in loops, each loop followed by a system-wide co-test to drive the optimization to next round.
	
	This approach has a great advantage in its simplicity of modeling: no new model is required beyond the domain-wide models that have been used in the traditional top-down design approach. Only a few insights about the coupling between constraints in different domains are demanded to supervise the loopback design. Conventional designing and optimization tools can also be directly applied in the design of each individual domain without any barrier. However, its disadvantages are also obvious to us: First, this loopback design iteratively enters the domains of control and communication in turns, so that cumulative expertise from both fields are required throughout the entire process of system design, which limits the flexibility of development cycle and keeps the development cost on a high level. Second, the system model used in this approach, especially about the way different layers are coupled to each other, lacks of abstract and generality. The resulted design is therefore highly dependent on specific system or application, and cannot be flexibly modified for demands such like system upgrade or scenario change.

	\subsection{Decomposing the system into inter-coupled modules}\label{subsec:coupled_modules}
	For a higher generality and flexibility, the co-design flow specified to certain application shall be decomposed into multiple abstracted domain-wide design stages that are independent from each other. More specifically, the application shall be modeled regarding a generic control mechanism, so that the application dependability can be formulated as a function of some control metric (or a vector of metrics). Similarly, the control model shall be set up with respect to a generic communication mechanism, characterizing the control metric (vector) as a function of some communication metric (vector). In the end, the application dependability model can be mapped with the both functions onto the communication model, and a stage-by-stage optimization can be therewith executed in a bottom-up fashion, in the inverse flow of this modular top-down modeling approach.
	
	By decomposing the cross-domain co-design process into several domain-specific design tasks that are only weakly coupled with each other, this approach allows a much more flexible development cycle to reduce the cost. Also, some conventional methods can still be easily adopted in the domain-specific design on each individual stage. Nevertheless, this paradigm shift strongly relies on a deep understanding of the cross-domain coupling. More specifically, a complete set of the coupling metrics that bridge the gaps between different domains has to be identified, and the metrics must be explicitly integrated to the analytical model of every domain. Unfortunately, our knowledge of such metrics is still insufficient, making this approach inapplicable in many applications. Moreover, since the model of every individual domain must be extended or modified to take account of the cross-domain couplings, some domain-specific conventional approaches will no more apply. For example, the Bode plot, which is one of the most convenient tool to analyze the stability of control system regarding consistent delay, is not more applicable when considering random delay caused by unreliable communications. In addition, some of such extended analytical models may be unavailable, which disables the application of analytical and numerical solvers. In this case, heuristic or data-driven solvers (including machine learning based methods) must be invoked for the optimization.

	\subsection{Identifying the coupling metrics}\label{subsec:coupling_metrics}
	The coupling metrics discussed in Sec.~\ref{subsec:coupled_modules} can only be identified with a detailed cross-domain error propagation model that we have mentioned in Sec.~\ref{subsec:err_prop}. Aiming at such a model, we first identify the error sources in NCS, and classify them into three categories w.r.t. their impact on the system's dependability, as listed in Tab.~\ref{tab:errors}.
	
	\begin{table}[!htpb]
		\centering
		\caption{Errors in NCS three categories}
		\label{tab:errors}
		\begin{tabular}{m{1.4cm}m{3.3cm}m{2.4cm}}
			\toprule[2px]
			\textbf{Category} 		& \textbf{Examples} 						& \textbf{Impact}\\
			\midrule[1.5px]
			value errors			& sensing error, computation error,
			source coding noise, quantization noise, 
			system disturbance, etc.
			& system state diverged from optimum\\\hline
			transmission errors		& retransmission, packet loss				& random information delay\\\hline
			application errors		& system outage, service failure			& reduced dependability on application layer\\
			\bottomrule[2px]
		\end{tabular}
	\end{table}
	
	Over the recent past years, some emerging metrics have been proposed to capture the propagation of errors through the three NCS layers, during which process the errors of different types are also transferred into each other. For example, the \emph{age of information} (AoI) concept~\cite{SKTM2019} in communication fully reflects the information delay, which is a critical parameter in the control domain. More precisely, peak AoI describes the information delay, while mean AoI captures both the mean and variance of the infomation delay~\cite{HZJ+2021}. There have been numerous studies such as \cite{FKB2020} reported to reveal how transmission errors will affect the AoI, wherewith the control domain can be well coupled to the communication domain~\cite{AVK+2019}. Further more, an variant of AoI, known as age of incorrect information (AoII), further takes value errors into account~\cite{MKAE2020}. Nevertheless, the state of the art is still insufficient to support a full modeling of such inter-domain coupling, especially when computing is also taken into account.
	
	\subsection{Generalized black-box model}\label{subsec:black_box}
	When deeply investigating the computation errors in the NCS model, it becomes more complicated, since computation errors can simultaneously impacting multiple domains. For example, a numerical algorithm in control domain that computes in an iterative fashion may be specified to different convergence thresholds, through which a balance between the computing delay and the computation error is taken. Since the computation and communication compete against each other for the shared time resource, a higher computing delay is leading to less time budget for communication, and therewith more transmission errors, i.e. degraded communication dependability. On the other hand, the reduction in computation error tends to reduce the value error, and therewith increase the control dependability. Similar phenomena have also been reported in other problems that can also occur in NCS, such as:
	\begin{itemize}
		\item \emph{Optimal pre-processing} where the communication dependability can be improved at a cost of higher computing latency;
		\item \emph{optimal learning} where the model accuracy can be improved for lower value error, but at a cost of higher observing latency; 
		\item \emph{optimal task offloading} where the communication error can be reduced by opportunistically using local computing instead of cloud computing, but at a cost of higher computing delay or higher value error.
	\end{itemize}
	These phenomena can make the coupling model between different domains highly complex, and the joint optimization problem probably analytically intractable.
	
	\begin{figure}[!htpb]
		\centering
		\includegraphics[width=.4\linewidth]{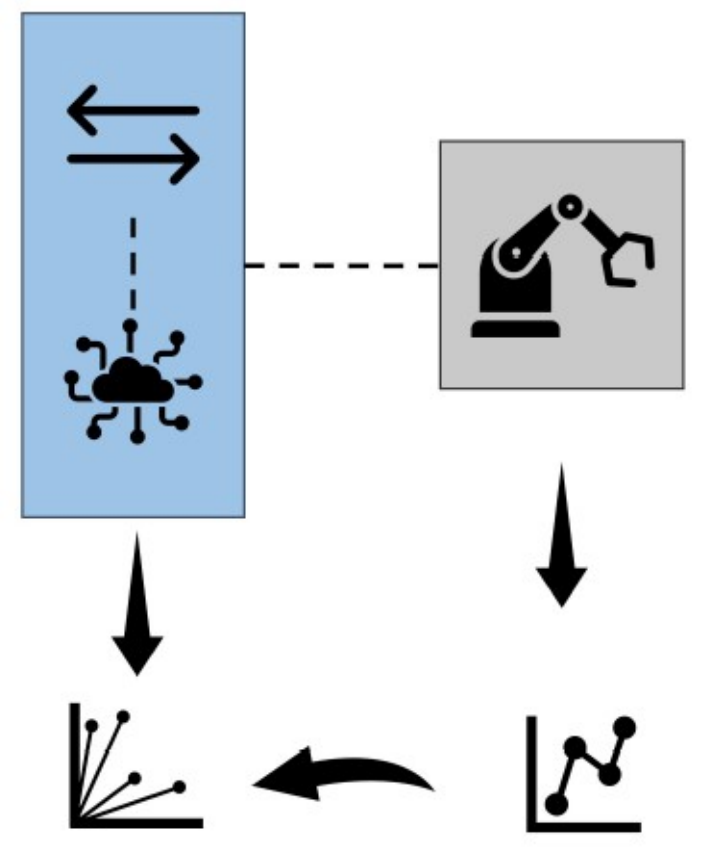}
		\caption{The proposed generalized black-box NCS model}
		\label{fig:blackbox}
	\end{figure}
	
	Towards an efficient solution to such problems, we propose to use a generalized black-box model for NCS and its dedicated CoCoCo optimization problem, as illustrated in Fig.~\ref{fig:blackbox}. The communication and control domains, together with their underlying computation algorithms, are jointly modeled as a packaged black-box that is independent from the physical implementation of application. This black-box module (the blue box top-left)  defines a generic interface to an abstracted application module (the gray module top-right), which is characterized by a \emph{generic error vector} defined in a multi-dimensional space of key performance indicators (KPIs). A certain implementation of the communication and control domains will define a feasibility region within this KPI space, which can be achieved by specifying the black-box. This feasibility region can be identified either theoretically, when a complete error propagation model is available, or empirically in a data-driven approach in the opposite case. On the other hand, for every specific application, a dependability model can be established to describe the application layer dependability as function of the generalized error vector. Thus, the CoCoCo problem to optimize the dependability can be considered as a mathematical programming problem, where the objective function if the application's dependability, and the search space is the feasibility region of the black-box.
	
	It is worth remarking that the classical top-down design approach actually restricts the search space into a (probably very small) subset of the feasibility region, and therefore fail to guarantee converging to the global optimum.
	
	\subsection{A multi-stage optimization formulation}\label{subsec:multi_stage}
	In this subsection, we convert the black-box model into a two-stage stochastic programming problem. We will illustrate the abstract models, but its tractability still needs to be exploited.
	
	\begin{mini!}[2]
		{z}{\mathbb{E}_{P_1}[Q_1(z,\omega_1)]}{\label{prob:outter}}{}
		\addConstraint{f(z)\leqslant 0\label{con:comm_model}}
		\addConstraint{\mathbb{E}_{P_1\times P_2}[Q_2(z,\omega_1,\omega_2)]\leqslant 1-R_{\text{sys,app}\_0}\label{con:sys_reliability}}
		\addConstraint{z\in\mathbb{R}^n\label{con:comm_params}}
	\end{mini!}

	\begin{mini!}[2]
		{z}{\mathbb{E}_{P_2}J(u)}{\label{prob:inner}}{[Q_1(z,\omega_1)]:=}
		\addConstraint{g(\dot{x},x,u,\omega_1,\omega_2)=0\label{ctrl_model}}
		\addConstraint{\{(x_0,t_0)\}=v(z,\omega_1)\label{con:init_state_set}}
		\addConstraint{u\in\mathbb{R}^m\label{con:ctrl_params}}
	\end{mini!}

	\begin{align}
		[Q_2(z,\omega_1,\omega_2)]:=&1-I(y)\label{eq:unreliab}\\
		y=&h(x,u,v(z,\omega_1),\omega_2)\label{eq:sys_state}\\
		I(y):=&\begin{cases}
			1,&y\in\mathcal{D}_\text{sys,app}\\
			0.&\text{o.w.}
		\end{cases}\label{eq:sys_reliability_indicator}
	\end{align}

\section{Applications}\label{sec:applications}
We envision the proposed black-box approach to revolutionize the paradigms in several critical industrial use scenarios, as listed below.
	
	\subsection{CoCoCo for dependable remote control}\label{subsec:dependable_remote_control}
	The most direct application is also the motivation of our proposal: the communication-control co-design for dependable remote control in NCS. As we have analyzed in Sec.~\ref{sec:approaches}, With decomposed modular design in communication and control domains, the flow of system design becomes more flexible and the development cost can be efficiently reduced. Also, with well identified coupling metrics, the generalized black-box model will be able to efficiently improve the optimization performance, by expanding the search space of system specification to the very extent of feasibility region.
	
	\subsection{Flexible manufacturing}\label{subsec:flexible_manufacturing}
	Classical factories generally have the control system embedded in the specialized manufacturing equipment, i.e. it is dedicated to the specific application. The control system is therefore also highly specified to this certain application, and cannot be flexibly reused for other applications (e.g. another product line). This architecture is significantly limiting the flexibility of manufacturing systems, bring huge cost to any replacement or upgrade of the production lines.
	
	For future manufacturing systems remote control, however, our proposed design framework is abstracting the application and decoupling it from the factory infrastructure. This allows to flexibly adapting the manufacturing system, including its control and communication modules, to different applications upon requirement, and therewith significantly reduces the capital expenditure (CAPEX) of flexible manufacturing.
	
	\subsection{Open manufacturing system}\label{subsec:open_system}
	Beyond the adaptation of manufacturing system upon different applications, the flexibility introduced by our proposed framework applies also to the different components within the factory infrastructure itself. Classical manufacturing systems generally have specialized modules of control, communication and computing, which are tuned to work with each other and integrated in a fixed manner. Specifications and calibrations to these modules are accomplished in prior to the delivery of the system, and provides little or no freedom of modular replacement. Since no information is available for the factories to estimate the application layer performance before purchasing, assembling, and practically testing the equipment, they cannot flexibly change or upgrade one single module without replacing the entire manufacturing system, even if it has a better option for that specific module, since the risk is too high to afford. On the other hand, the module vendors can hardly sell their products directly to the end users (i.e. the factories) without intervention of the system vendors, and therefore bonded to the latter. Thus, in the current business model, the system vendors are holding unfair advantages over the other two stakeholders.
	
	Our proposed approach, however, has a rich potential in modular modeling of manufacturing systems. The factories will be able to accurately estimate the achievable performance of a customized system, which will not only significantly reduce the CAPEX of modifying or upgrading production lines, but also open the gate of this huge market for direct entrance of module vendors. This openness of manufacturing system design will certainly lead to a more fair, competitive, and free market in this industry, as OpenRAN has already done to the wireless network industry~\cite{KY2002}.
 	
	\subsection{SLA design for business applications}\label{subsec:SLA}
	In previous, we have mostly discussed about the co-design of the data links of communication systems with the setting of controllers in control systems. In this last subsection, we will also mention about the potential impact of CoCoCo research on the system-level design for the private networks implemented on industrial sectors under future 6G-to-business models. 
 
 In the era of 5G, industrial enterprises have already shown a strong interests of URLLC on industrial control scenarios that are highly inconvenient for wired networks, e.g., the Industrial Ethernet or CAN protocols. However, it is still hard to call the promotion of URLLC to business applications as a huge success. For industrial sectors, the core concern is to maintain a “ultra-reliable low E2E time delay” communication network 24hrs-7ds with a reasonable total budget including the construction, operation \& maintenance cost. In a SLA contract between industrial enterprises and the wireless communication service providers, it often asks for the guarantee of performance reliability, e.g., the E2E time-delay reliability or data rate reliability. Thus, it also arises the following open problems: 
 \begin{enumerate} 
 \setlength{\itemsep}{-2ex}  
 \setlength{\parskip}{0ex} 
 \setlength{\parsep}{0ex}
\item The design of quantitative reliability indicators for SLA contracts: for instance, to guarantee a given performance reliability of industrial applications, what level of reliability for E2E latency should be ensured?\hfil\break
\item The joint redundancy allocation of industrial \& communication systems: for instance, the type, model, location \& number of redundancies of industrial \& communication equipment, such as the machine tools (in manufacturing systems) \& the AAU, BBU, RRU in the RAN (of communication networks), should be co-designed to maximize the quality of wireless communication. \hfil\break
\item The maintenance of communication networks: for instance, can we design proper predictive maintenance schedules, to minimize the risk of reliability decline, due to the failures or inappropriate parameter settings that appeared in the communication networks?\hfil\break
\end{enumerate}
We have mentioned of problem 1) at equation \ref{1} in section \ref{subsec:depend_ind}. For the rest of the problems, we conclude them as system level CoCoCo problems: an additional outer-stage decision that determines the system structure should be added to the previous two-stage stochastic programming problem that presented in section \ref{subsec:multi_stage}.

	\section{Conclusion}\label{sec:conclusion}
	In this paper, we have studied the dependability of I4.0 applications in the upcoming 6G era, which are calling for novel solutions beyond xURLLC to fulfill their performance requirements. Starting with the dependability model of networked control systems, we have analyzed the challenges in industrial dependability, and proposed a novel framework of communication-control co-design by evolving the conventional top-down design approach step-by-step. Besides achieving dependable remote control, our proposal also exhibits rich potential in enabling flexible manufacturing and shifting the business paradigm of future industry, which we have also briefly envisaged.

	\section*{Acknowledgement}
	This work has been partly funded by the European Commission through the H2020 project Hexa-X (Grant Agreement no. 101015956).
	
	\ifCLASSOPTIONcaptionsoff
	\newpage
	\fi
	

\end{document}